\documentclass[prb,twocolumn,amsmath,amssymb,showpacs,superscriptaddress]{revtex4}
\usepackage{amsmath}
\usepackage{graphicx}
\usepackage{dcolumn}
\usepackage{bm}
\def\be{\begin{eqnarray}}
\def\ee{\end{eqnarray}}
\begin{document}
\title{Fluctuations in superconducting rings with two order parameters}

\author{Jorge Berger}
\affiliation{Department of Physics and Optical Engineering, Ort-Braude College, P.O. Box 78,
21982 Karmiel, Israel}
\author{Milorad V. Milo\v{s}evi\'c}
\affiliation{Departement Fysica, Universiteit Antwerpen,
Groenenborgerlaan 171, B-2020 Antwerpen, Belgium}

\begin{abstract}
Starting from the Ginzburg--Landau energy functional, we discuss how the
presence of two order parameters and the coupling between them influence a
superconducting ring in the fluctuative regime. Our method is exact, but requires numerical implementation.
We also study approximations for which some analytic expressions can be obtained, and check their ranges of validity.
We provide estimates for the temperature ranges where
fluctuations are important, calculate the persistent current in
MgB$_2$ rings as a function of temperature and enclosed flux, and point out its
additional dependence on the cross-section area of the ring.
We find temperature regions in which fluctuations enhance the persistent currents and regions where they inhibit the persistent current.
The presence of two order parameters that can fluctuate independently always leads to larger averages of the order parameters at $T_c$, but only for appropriate parameters this yields larger persistent current.
In cases of very different material parameters
for the two coupled condensates, the persistent current is inhibited.
\end{abstract}


\maketitle

\section{Introduction}
Fluctuations are extremely important near phase transitions, and
have therefore been the subject of intense research in the past.
Particularly in superconductivity, it has been shown that thermally
driven electronic fluctuations, i.e. formation and dissociation of
Cooper pairs close to the critical temperature $T_c$, can affect all
relevant properties of a superconductor.\cite{rev} 
Techniques that incorporate thermal fluctuations to the Ginzburg--Landau model are described in a recent review.\cite{Baruch}
Fluctuations in
mesoscopic loops are particularly interesting because their
critical temperature is reduced in an oscillatory fashion as a
function of the magnetic field---a phenomenon known as the Little--Parks (LP)
effect.\cite{lp}
More importantly, as LP oscillations are directly
related to flux (vorticity) entry in superconductors, one can identify
the magnetic fields for which fluctuations are particularly
important, as is the case of half-integer flux
values.\cite{koshnick} The latter experiment\cite{koshnick} detected current in the ring above $T_c$, a
clear signature of fluctuations, and may be regarded as a paradigm for thin superconducting ring behavior, a case for which the 
theory for thermal fluctuations is known exactly.\cite{vOR}
The additional influence of quantum fluctuations was addressed in Ref.~\onlinecite{scoreg}.

Superconductivity is essentially a macroscopic quantum state with
long-range phase coherence, therefore described as a single wave
function. Superconductors with several order
parameters have recently attracted great attention due to the discovery of
MgB$_2$ and high $T_c$ superconductivity in pnictides. In such
cases, thermal excitation allows contributions from multiple wave
functions and one may expect a dramatically different behavior of the
system. With that as motivation, we here explore the interplay of
the wave functions and thermal fluctuations in superconducting rings
with two order parameters. The rings we will consider need not be
made of a two-band superconductor, but may also consist of two thin
superimposed superconducting rings,\cite{bluhm} possibly separated
by an isolating layer, such as the active part in readily made
experiments with annular Josephson junctions.\cite{Monaco}

We conduct our theoretical analysis in the framework of the
Ginzburg--Landau (GL) theory. The multiband GL 
equations were developed long ago;\cite{Geil} in the case of two bands the free energy
density has the form \be f=\sum_{\nu =1,2}\left(\tilde{a}_\nu
|\tilde{\Delta }_\nu |^2+\frac{\tilde{b}_\nu}{2}|\tilde{\Delta }_\nu
|^4+ \tilde{K}_\nu |{\bf \Pi} \tilde{\Delta }_\nu
|^2\right)\nonumber\\-\tilde{\gamma} (\tilde{\Delta }_1
\tilde{\Delta }_2^*+\tilde{\Delta }_2 \tilde{\Delta }_1^*)\;,
\label{def} \ee where $\tilde{\Delta }_{1,2}$ are the order
parameters, $\tilde{a }_{1,2}$, $\tilde{b}_{1,2}$, $\tilde{K
}_{1,2}$ and $\tilde{\gamma}$ are material parameters and ${\bf
\Pi}={\bm \nabla} +2\pi i {\bf A}/\Phi_0$, with ${\bf A}$ the vector potential and $\Phi_0$ the
superconducting flux-quantum. Zhitomirsky and Dao\cite{ZD} obtained
expressions for the material parameters in a multiband
superconductor using Gor'kov's technique. 
Kogan and Schmalian\cite{KSch} recently emphasized that consistency imposes conditions on the temperature dependence of these coefficients, which results in the same coherence length for both order parameters in a two-band superconductor. Shanenko {\it et al.}\cite{shan} went on to show the importance of terms of higher order in temperature, and the resulting  separation of characteristic lengths for the two bands. We should note, however, that fluctuations move the order parameters astray from equilibrium, so that in general their ratio is not constant. Moreover, besides two-band superconductors, we are interested in relating our results to additional systems, where the coefficients in Eq.~(\ref{def}) may have a different temperature dependence. In this paper we thus adopt the standard GL approach, where the material parameters in Eq.~(\ref{def}) are arbitrary functions of the temperature and any required restriction will be a particular case.

\section{Method}
In this paper we deal with one dimensional superconducting rings
with two order parameters, extending the results
obtained by von Oppen and Riedel\cite{vOR} from single to
two order parameters. Instead of the formalism of
Ref.~\onlinecite{vOR}, we will follow a slightly different approach,
which in our view is conceptually simpler. We start by absorbing the
coefficients $\tilde{K }_{1,2}$ into the order parameters and by
switching to a gauge invariant formulation, i.e.\ we define \be
\Delta_\nu (\theta )=\exp\left(\frac{2\pi i R}{\Phi_0}\int_0^\theta
A(\theta ')d\theta '\right)\sqrt{\tilde{K}_\nu}\tilde{\Delta }_\nu
(\theta ) \;, \label{gauge} \ee where $R$ is the radius of the ring,
$\theta $ is the angle along the ring and $A$ is the tangential
component of ${\bf A}$. Likewise we define \be a_\nu
=\frac{\tilde{a}_\nu}{\tilde{K}_\nu}  \;, \;\;\; b_\nu
=\frac{\tilde{b}_\nu}{\tilde{K}_\nu^2}  \;, \;\;\; \gamma
=\frac{\tilde{\gamma}}{\sqrt{\tilde{K}_1 \tilde{K}_2}}\;.
\label{redef} \ee

With these definitions the free energy density becomes \be
f=\sum_{\nu =1,2}\left(a_\nu |\Delta _\nu
|^2+\frac{b_\nu}{2}|\Delta_\nu |^4+\frac{1}{R^2} \left|\frac{d\Delta
_\nu}{d\theta }\right|^2\right)\nonumber\\-\gamma (\Delta_1
\Delta_2^*+\Delta_2 \Delta_1^*)\;. \label{f} \ee With the
normalizations we are using, $a_\nu^{-1/2}$ is the coherence length
for the order parameter $\Delta_\nu$ in the absence of coupling to
the other order parameter. As in Ref.~\onlinecite{vOR}, we consider
a uniform 1D ring, which in particular does not alter the applied
field (screening is negligible), so that no magnetic energy has to
be added to $f$.

In order to have a more intuitive picture of the problem, we
represent the complex order parameters $\Delta_\nu$ by
two-dimensional real vectors ${\bf r}_\nu$, such that in polar
coordinates $r_\nu =|\Delta_\nu |$ and the angular coordinate
$\vartheta_\nu$ is the phase of $\Delta_\nu$. Integrating $f$ over
the volume of the ring, its free energy becomes \be
F=\frac{2w}{R}\int_{-\pi}^\pi d\theta \left( \frac{1}{2}\left|
\frac{d{\bf r}_1}{d\theta }\right|^2+  \frac{1}{2}\left|\frac{d{\bf
r}_2}{d\theta }\right|^2 +V\right) \;, \label{F} \ee where $w$ is
the cross section of the ring and \be
V=\frac{R^2}{2}\left(a_1r_1^2+a_2r_2^2+\frac{b_1}{2}r_1^4+\frac{b_2}{2}r_2^4\right.\nonumber\\
\left.-\frac{}{}2\gamma r_1r_2\cos (\vartheta _1-\vartheta
_2)\right)\;. \label{V0} \ee As in Ref.~\onlinecite{vOR}, ${\bf
r}_{1,2} (\theta )$ may be regarded as the trajectories of two
fictitious particles during a period of time $-\pi\le\theta \le\pi$.
The first two terms in the integrand of Eq.~(\ref{F}) then represent
their kinetic energy and $V$ their potential energy.

Following Ref.~\onlinecite{Scal}, a pair of functions ${\bf r}_\nu
(\theta )$ is interpreted as a microstate of the system and $F$ as
the energy of the system for that microstate. It follows that up to
an irrelevant multiplicative constant the partition function is \be
Z=\int {\cal D}{\bf r}_1{\cal D}{\bf r}_2\exp (-F/k_B T) \;,
\label{Z1} \ee where $T$ is the temperature and $\int {\cal D}{\bf
r}_1{\cal D}{\bf r}_2$ denotes integration over all functions ${\bf
r}_\nu (\theta )$ with appropriate periodicity. Since $\tilde{\Delta
}_\nu$ are single valued, $r_\nu (\theta =\pi )=r_\nu (\theta =-\pi
)$ and  $\vartheta _\nu (\theta =\pi )=\vartheta _\nu (\theta =-\pi
)+2\pi\varphi $, where $\varphi $ is the flux enclosed by the ring
divided by $\Phi_0$.

Using slight adaptations of Eqs.\ (2.14), (2.15), (2.16), (2.22) and
(2.23) in Ref.\ \onlinecite{Scal}, $Z$ can be brought to the form
\be Z=&&\sum_n\exp(-2\pi\varepsilon_n/S)\nonumber\\ &&\times\int
d{\bf r}_1d{\bf r}_2\Psi_n^*[{\bf r}_\nu (\theta =\pi)]\Psi_n[{\bf
r}_\nu (\theta =-\pi)] \;. \label{Z2} \ee Here $\varepsilon_n$ and
$\Psi_n[{\bf r}_\nu ]$ are a complete set of eigenvalues and
normalized eigenfunctions of the fictitious Hamiltonian \be
H=-\frac{S^2}{2}\left( \nabla_1^2+\nabla_2^2\right)+V \;, \label{H0}
\ee where the Laplacian $\nabla_\nu^2$ acts on ${\bf r}_\nu$ and
$S=k_BTR/2w$. $S$ has dimensions of surface in the plane of the
trajectories of ${\bf r}_{1,2}$ and dimensions of force in reality.
The integral in Eq.~(\ref{Z2}) is taken over the entire planes of
motion for each particle, but for every argument ${\bf r}_\nu$ in
$\Psi_n$ we have to take the corresponding argument in $\Psi_n^*$.

We note now that the angular momentum operator $L_z=-i(\partial
/\partial \vartheta _1+\partial /\partial \vartheta _2)$ commutes
with $H$. We can therefore choose the set of eigenfunctions
$\{\Psi_n\}$ with well defined angular momentum, i.e.\ they can obey
$L_z\Psi_{n,\ell}=\ell\Psi_{n,\ell}$ and therefore have the form
$\Psi_{n,\ell}({\bf r}_1,{\bf
r}_2)=\sum_{\ell_1}\psi_{n,\ell,\ell_1}
(r_1,r_2,\vartheta_1-\vartheta_2)\exp [i(\ell_1\vartheta_1+(\ell
-\ell_1)\vartheta_2)]$, with $\ell$ and $\ell_1$ integers. In view
of the periodicity conditions of ${\bf r}_\nu$, it follows
$\Psi_{n,\ell}[{\bf r}_\nu (\theta =\pi)]=\exp (2\pi i \ell\varphi
)\Psi_{n,\ell}[{\bf r}_\nu (\theta =-\pi)]$. By substitution of this
result into Eq.~(\ref{Z2}) we obtain \be Z=\sum_\ell\exp (-2\pi i
\ell\varphi )Z_\ell \;, \label{Z3} \ee with \be Z_\ell
=\sum_n\exp(-2\pi\varepsilon_{n,\ell}/S) \;, \label{ZL} \ee where
summation in Eq.~(\ref{Z3}) is made over all integers and in
Eq.~(\ref{ZL}) over all the states with total angular momentum
$\ell$. Since $H$ is invariant under the transformation $\{\vartheta
_1,\vartheta _2\}\rightarrow \{-\vartheta _1,-\vartheta _2\}$,
$\varepsilon_{n,-\ell}=\varepsilon_{n,\ell}$ and we can also write
\be Z=Z_0+2\sum_{\ell =1}^\infty\cos (2\pi\ell\varphi )Z_\ell \;.
\label{Z4} \ee

Once the partition function is known, all the equilibrium quantities
can be derived from it. The average current around the ring is \be
\langle I\rangle =\frac{k_BT}{\Phi_0}\frac{\partial \ln Z}{\partial
\varphi }=-\frac{k_BT}{\Phi_0}\frac{4\pi}{Z}\sum_{\ell
=1}^\infty\sin (2\pi\ell\varphi )\ell Z_\ell \;. \label{current} \ee
Following similar steps to those that led to Eq.~(\ref{Z4}), we
obtain that the statistical average of any function of the absolute
values of the order parameters is \be \langle
p(|\Delta_1|,|\Delta_2|)\rangle =\frac{1}{Z}\left[ P_0+2\sum_{\ell
=1}^\infty\cos (2\pi\ell\varphi )P_\ell \right] \,, \label{anyp} \ee
where $P_\ell =\sum_n\exp(-2\pi\varepsilon_{n,\ell}/S)\langle n\ell
|p(r_1,r_2)|n\ell\rangle$. Here we introduced the matrix element
$\langle n\ell |p(r_1,r_2)|n\ell\rangle =\int d{\bf r}_1d{\bf
r}_2p(r_1,r_2)|\Psi_{n,\ell}({\bf r}_\nu) |^2$, which may in
practice be evaluated in any convenient basis.

\section{Evaluations}
\subsection{High temperature}

Let $T$ be sufficiently higher than $T_c$, so
that the order parameters are small and the quartic terms in $V$ can
be neglected. For high $T$ we can also assume
$a_1+a_2>\sqrt{(a_1-a_2)^2+4\gamma ^2}$ and define the quantities
\be \eta =\frac{a_1-a_2}{2\sqrt{(a_1-a_2)^2+4\gamma ^2}} \;,\nonumber\\
\xi_{3,4}^2=\frac{2}{a_1+a_2\mp\sqrt{(a_1-a_2)^2+4\gamma ^2}} \;.
\label{constants} \ee For $\gamma =0$ and $a_1>a_2$,
$\xi_{3,4}=a_{2,1}^{-1/2}$; we therefore regard $\xi_{3,4}$ as a
sort of coherence lengths in the presence of coupling.

We can now define a rotation in the 4D space of both particles
through \be {\bf r}_1=\sqrt{\frac{1}{2}-\eta}\;{\bf
r}_3-\sqrt{\frac{1}{2}+\eta}\;{\bf r}_4 \;,\nonumber\\  {\bf
r}_2=\sqrt{\frac{1}{2}+\eta}\;{\bf
r}_3+\sqrt{\frac{1}{2}-\eta}\;{\bf r}_4 \;. \label{rotation} \ee
With this transformation, the ``potential energy" in Eq.~(\ref{V0})
becomes \be V_{\rm quad}=(R^2/2)[ (r_3/\xi_3)^2+(r_4/\xi_4)^2] \;.
\label{Vquad} \ee

The following features should be noted: (i) in the coordinates
$\{{\bf r}_3,{\bf r}_4\}$, $\nabla_1^2+\nabla_2^2$ still has the
meaning of the Laplacian in the 4D space; (ii) since on passing from
$\theta =-\pi$ to $\theta =\pi$ both $\vartheta _1$ and $\vartheta
_2$ increase by $2\pi\varphi $, and since ${\bf r}_{3,4}$ are linear
combinations of ${\bf r}_{1,2}$ with fixed coefficients, also their
angles $\vartheta _{3,4}$ increase by $2\pi\varphi $; (iii) the
angular momentum operators $-i\partial /\partial \vartheta _{3,4}$
for each separate particle now both commute with the Hamiltonian;
(iv) the total angular momentum equals the sum of the angular
momenta of each particle and the total energy equals the sum of
their energies. As a consequence of these features, the partition
function in Eq.~(\ref{Z3}) becomes $Z=Z^{(3)}Z^{(4)}$, where
$Z^{(\nu)}$ is the value of $Z$ obtained when $\varepsilon_{n,\ell}$
is replaced with the value that corresponds to particle $\nu$ only.
It follows that equilibrium quantities such as the average current
or the average energy will equal the sum of the separate
contributions of particles 3 and 4.

The Hamiltonian of the fictitious particles 3 and 4 is just that of
two decoupled harmonic oscillators and its eigenvalues are well
known: $\varepsilon_{n,\ell}^{(3,4)}=(SR/\xi_{3,4})(2n+|\ell|+1)$.
The sum in Eq.~(\ref{ZL}) becomes a geometric series and we obtain
\be Z_\ell^{(3,4)}=\frac{e^{-2\pi|\ell |R/\xi_{3,4}}}{2\sinh2\pi
R/\xi_{3,4}} \;. \label{ZL34} \ee Substitution into Eq.~(\ref{Z3})
gives \be Z^{(3,4)}=[2(\cosh 2\pi R/\xi_{3,4}-\cos 2\pi\varphi
)]^{-1} \label{Z34} \ee and the average current in the ring equals
\be &&\langle I_{\rm quad}\rangle=\nonumber\\&&-(2\pi\sin
2\pi\varphi\,
k_BT/\Phi_0)\left[(\cosh 2\pi R/\xi_3-\cos 2\pi\varphi )^{-1}\right.\nonumber\\
&&\left.+(\cosh 2\pi R/\xi_4-\cos 2\pi\varphi )^{-1}\right] \;.
\label{Iquad} \ee

From Eq.~(\ref{Iquad}) we can obtain the Little--Parks temperature,
i.e., the temperature for the onset of superconductivity in the
absence of fluctuations. Without fluctuations the current vanishes
above this onset; this is implemented by taking the limit $k_BT\rightarrow 0$ in the first factor in Eq.~(\ref{Iquad}). At the LP temperature the current becomes nonzero, requiring divergence of the second factor, i.e. $iR/\xi_3=\varphi\mod 1 $. From here, the
LP condition is \be R^2[\sqrt{(a_1-a_2)^2+4\gamma
^2}-(a_1+a_2)]=2\varphi ^2 \,, \label{LP} \ee where for simplicity
of notation we restrict ourselves to the range $|\varphi |\le 1/2$.
Near $T_c$ we can write $a_\nu =a_{\nu c}-\alpha_\nu\tau$, with
$\tau =(T_c-T)/T_c$ and $a_{1c}a_{2c}=\gamma^2 $; if in addition
$(\varphi /R)^2\ll\alpha_{1,2}$, condition \eqref{LP} is fulfilled
for \be \tau =\tau_{\rm LP}=\frac{a_{1c}+a_{2c}}{a_{1c}\alpha
_2+a_{2c}\alpha _1}\frac{\varphi^2}{R^2} \,. \label{tauLP} \ee

\subsection{Hartree approximation}
In the potential given by Eq.~(\ref{V0}) we now make the replacement
$(b_{1,2}/2)r_{1,2}^4\rightarrow b_{1,2}\langle r_{1,2}^2\rangle
r_{1,2}^2$. At high temperatures both terms are negligible and at
low temperatures, where fluctuations can be neglected, they both lead
to the same ``force" $-{\bm \nabla}V$.\footnote{This is due to the
fact that in the absence of fluctuations thermodynamic quantities
are determined by the derivatives of $F$ with respect to
$r_{1,2}^2$, which in one case give $b_{1,2}r_{1,2}^2$ and in the
other, the equivalent $b_{1,2}\langle r_{1,2}^2\rangle$.}

With this approximation the potential becomes again quadratic, so
that we can still use the results of the previous section by
substituting $\eta \rightarrow \eta '$ and $\xi_{3,4}\rightarrow
\xi_{3,4}'$, with \be \eta
'=\frac{a_1'-a_2'}{2\sqrt{(a_1'-a_2')^2+4\gamma ^2}} \;,\nonumber\\
\xi_{3,4}^{\prime
2}=\frac{2}{a_1'+a_2'\mp\sqrt{(a_1'-a_2')^2+4\gamma ^2}}
\;,\nonumber\\ a_{1,2}'=a_{1,2}+b_{1,2}\langle r_{1,2}^2\rangle \;.
\label{constantsprime} \ee

In order to implement this approximation, we have to evaluate
$\langle r_{1,2}^2\rangle$. $\langle r_{3,4}^2\rangle$ are given by
\be \langle r_{3,4}^2\rangle &&=-\frac{S\xi_{3,4}'}{2\pi
R^2}\frac{\partial \ln Z^{(3,4)}}{\partial
(1/\xi_{3,4}')}\nonumber\\&&=\frac{S\xi_{3,4}'}{R} \frac{\sinh 2\pi
R/\xi_{3,4}'}{\cosh 2\pi R/\xi_{3,4}'-\cos 2\pi\varphi } \; ;
\label{avr34} \ee on the other hand, since $\langle {\bf r}_3\cdot
{\bf r}_4\rangle =0$, $\langle r_{1,2}^2\rangle =(1/2\mp \eta')
\langle r_3^2\rangle +(1/2\pm \eta') \langle r_4^2\rangle$, hence
$\langle r_{3,4}^2\rangle =(1/2\mp 1/4\eta')\langle
r_1^2\rangle+(1/2\pm 1/4\eta')\langle r_2^2\rangle$. Substituting this into
Eq.~(\ref{avr34}) leads to \be &&\left(\frac{1}{2}\mp
\frac{1}{4\eta'}\right)\langle r_1^2\rangle+\left(\frac{1}{2}\pm
\frac{1}{4\eta'}\right)\langle
r_2^2\rangle=\nonumber\\&&\frac{S\xi_{3,4}'}{R} \frac{\sinh 2\pi
R/\xi_{3,4}'}{\cosh 2\pi R/\xi_{3,4}'-\cos 2\pi\varphi } \; .
\label{2eqs} \ee This is a system of two equations for obtaining
$\langle r_1^2\rangle$ and $\langle r_2^2\rangle$.

For a general situation, Eqs.~(\ref{2eqs}) have to be solved
numerically, but we can find asymptotic expressions for some special
situations of interest. Sufficiently above the critical temperature
$T_c$ one usually has $2\pi R/\xi_{3,4}'\gg 1$, so that the
fractions at the right of Eqs.~(\ref{2eqs}) reduce to 1. Near $T_c$
and for a range in which $b_\nu \langle r_\nu^2\rangle\ll\alpha _\nu
|\tau |\ll a_\nu$ we obtain 
\be &&
\langle |\Delta
_{1,2}|^2(T)\rangle =\langle r_{1,2}^2\rangle
\approx\nonumber\\&&\frac{k_BT}{2(a_{1c}+a_{2c})^{3/2}
w}\left(a_{1c,2c}+
\frac{a_{2c,1c}(a_{1c}+a_{2c})}{\sqrt{(a_{1c}\alpha _2+a_{2c}\alpha
_1)|\tau |}}\right) \;. \label{largetau} \ee 
It is interesting to
note that $\langle |\Delta _{1,2}|^2(T)\rangle$ decreases very
moderately with $T-T_c$. If the radius of the ring is sufficiently
large, we still have $2\pi R/\xi_{3,4}'\gg 1$ at $T=T_c$; assuming
that $b_\nu \langle r_\nu^2\rangle\ll a_\nu$ still holds leads to
\be &&\langle
|\Delta_{1,2}|^2(T_c)\rangle\approx\left(\frac{a_{2c,1c}^2k_B^2T_c^2}{4(a_{1c}+a_{2c})\beta_{1,2}w^2}\right)^{1/3}\nonumber\\&&+
\frac{a_{1c,2c}[3\beta_{1,2}-a_{2c,1c}(b_1+b_2)]k_BT_c}{6(a_{1c}+a_{2c})^{3/2}\beta_{1,2}
w} \;, \label{largeR} \ee with $\beta
_{1,2}=b_{1,2}a_{2c,1c}+b_{2,1}a_{1c,2c}^3/\gamma ^2$. Comparison of
Eqs.~(\ref{largetau}) and (\ref{largeR}) with the values of $|\Delta
_{1,2}|^2$ in the absence of fluctuations enables us to estimate the range
of temperatures for which fluctuations are important.

\begin{figure}
\includegraphics[width=\linewidth]{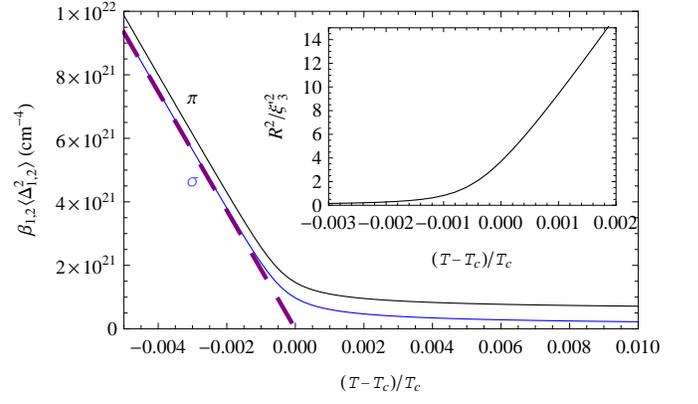}
\caption{\label{Delta12}Average Cooper-pair densities $|\Delta
_{\nu=1,2(=\sigma ,\pi )}|^2$ in the ring as a function of
temperature, calculated in the Hartree approximation. Each $|\Delta
_\nu|^2$ has been multiplied by $\beta _\nu$ for the purpose of
comparison with the values in the absence of thermal fluctuations
(dashed line). The sample is a ring of radius $10^{-4}\,$cm and
cross section $10^{-10}\,$cm$^2$ that encloses no magnetic flux. The
material parameters are those of MgB$_2$, taken from
Ref.~\onlinecite{Golubov}. Inset: $(R/\xi'_3)^2$ for the same ring.}
\end{figure}

Figure \ref{Delta12} shows the values of $\langle \Delta
_{1,2}^2\rangle =\langle r_{1,2}^2\rangle$ near $T_c$ for MgB$_2$,
using the material parameters reported in Ref.~\onlinecite{Golubov}.
In the absence of thermal fluctuations, $\beta _{1,2}\langle \Delta
_{1,2}^2\rangle$ is given by the dashed line for both bands. The
inset in the figure shows the behavior of $R/\xi'_3$ near $T_c$;
$\xi'_4\ll\xi'_3$ remains practically constant in this range.

\subsection{Exact evaluation}
We define the basis Hamiltonian \be
H_B(k_1,k_2)=-\frac{S^2}{2}\left(
\nabla_1^2+\nabla_2^2\right)+\frac{R^2}{2}(k_1^2r_1^2+k_2^2r_2^2)
\;, \label{basis} \ee which has the known set of eigenfunctions 
\be
&&\psi_{n_1,\ell_1,n_2,\ell_2}({\bf r}_1,{\bf r}_2)=C\prod_{\nu
=1,2}r^{|\ell_\nu |}e^{-Rk_\nu
r_\nu^2/2S}\,\nonumber\\
&&~~~~~~~\times\;_1F_1(-n_\nu ,|\ell_\nu |+1,Rk_\nu
r_\nu^2/S)e^{i\ell_\nu\vartheta} \; \label{eigf} 
\ee with
eigenvalues $RS[k_1(2n_1+|\ell_1|+1)+k_2(2n_2+|\ell_2|+1)]$. Here
$C$ is the normalization constant and $_1F_1$ is Kummer's
hypergeometric function. We can then evaluate any matrix element
$H_{i,j}=\langle\psi_i |H|\psi_j\rangle$, where the functions
$\psi_{i=1,\dots,N}$ are the functions with lowest eigenvalues within
the basis of the Hilbert space provided by Eq.~(\ref{eigf}). More
precisely, when evaluating $Z_\ell$, we include in the set
$\{\psi_i\}$ only eigenfunctions that obey $\ell_1+\ell_2=\ell$. If
$N$ is sufficiently large, then the lowest eigenvalues of the matrix
$(H_{i,j})$ will be a sufficiently accurate approximation for the
lowest eigenvalues of the operator $H$. In practice, rather than
fixing $N$, we fix maximum values for $n_1$, $|\ell_1|$, $n_2$ and
$|\ell_2|$.

We are interested in choosing $k_1$ and $k_2$ so that an accurate
approximation is obtained without $N$ becoming prohibitively large.
From Eq.~(\ref{eigf}) we see that the scale of $\langle
r_\nu^2\rangle$ is given by $S/Rk_\nu$; we therefore set $k_\nu
=pS/R\langle r_\nu^2\rangle$, where $\langle r_\nu^2\rangle$ is
obtained from the Hartree approximation and $p$ is still a free
parameter. Since for a good approximation the eigenvalues should
actually be independent of $p$, we mimic this situation by
minimizing the lowest eigenvalue of $(H_{i,j})$ with respect to $p$.

\begin{figure}
\includegraphics[width=\linewidth]{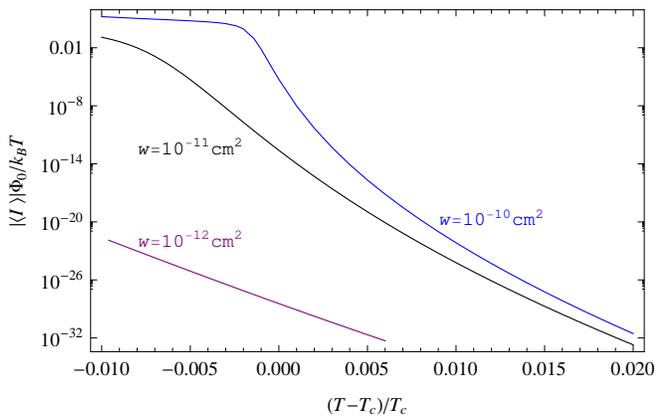}
\caption{\label{Ioftw} Average current in a two-order-parameter
superconducting ring as a function of temperature. For the top curve
the parameters are as in Fig.~\ref{Delta12}, except that the
normalized flux is $\varphi =0.25$. For the lower curves the cross
section $w$ of the ring is smaller. The current was evaluated using
a truncated basis of the Hilbert space, provided by the functions in
Eq.~(\ref{eigf}) with maximum quantum numbers $n_{1,{\rm max}}=11$,
$|\ell_1|_{\rm max}=22$, $n_{2,{\rm max}}=4$, and $|\ell_2|_{\rm
max}=8$.}
\end{figure}

Figure \ref{Ioftw} shows the currents as functions of the
temperature obtained with this method for rings of three different
cross sections, with the material parameters of MgB$_2$. For
completeness, the figure includes also values of current that are
too small to be experimentally observable. The temperature range
covered here is much wider than the range presented in
Ref.~\onlinecite{vOR}, where the temperature scale is given by the
Thouless correlation energy (divided by $k_B$) which is of the order
of the LP temperature; for a MgB$_2$ ring with radius of the order of a micron the LP temperature is of the
order of $10^{-5}\,T_c$.

\begin{figure}
\includegraphics[width=\linewidth]{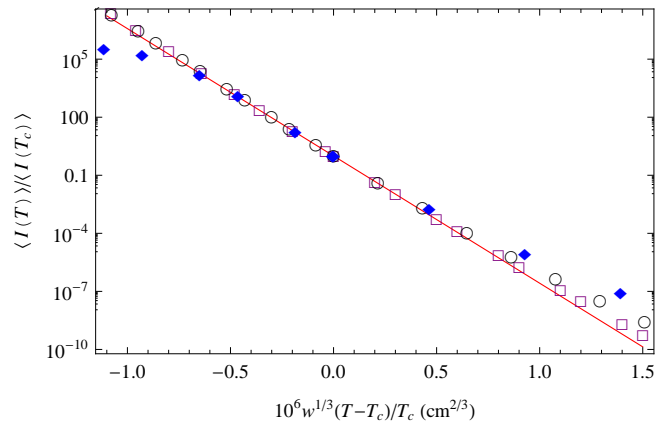}
\caption{\label{universalTc} Scaling of the function  $\langle
I(T)\rangle$ with the cross section. The parameters are the same as
in Fig.~\ref{Ioftw}. $\blacklozenge$: $w=10^{-10}\,$cm$^2$;
$\bigcirc$: $w=10^{-11}\,$cm$^2$; $\square$: $w=10^{-12}\,$cm$^2$.
The straight line is a guide for the eye.}
\end{figure}

For sufficiently high temperature (the required temperature
decreases with cross section $w$), the current becomes independent
of the cross section. For the parameters taken here and $T=3T_c$, a
ring with $w=10^{-10}{\rm cm}^2$ and a ring with $w=10^{-8}{\rm
cm}^2$ carry the same average current (within 1\% difference). At
the other extreme, far below $T_c$, i.e.\ where fluctuations are
unimportant, $\langle I\rangle\propto w$. However, the 
dependence of $\langle I\rangle$ on $w$ near $T_c$ is not
intermediate: we notice in Fig.~\ref{Ioftw} that decrease of $w$ by an
order of magnitude near $T_c$ leads to a current decrease of several
orders of magnitude, whereas for intermediate behavior the current would decrease by one order of magnitude at most. Figure \ref{universalTc} shows the scaled
current against the scaled temperature for the same rings as in
Fig.~\ref{Ioftw}, close to $T_c$. We notice that, in spite of the
moderate influence of temperature on $\langle |\Delta
_{1,2}|^2\rangle$ predicted by Eq.~(\ref{largetau}), the current
decreases exponentially. We also find that for smaller cross
sections the rate of change of $\langle I\rangle$ is slower. We
empirically found that the scaling $w^{1/3}$ leads to a
universal curve.

We attribute didactic interest to understanding the behavior of our
methods far below $T_c$. There, convergence of the series in Eqs.\
(\ref{Z4}) and (\ref{current}) becomes slow, and numeric
implementation of the exact evaluation becomes inefficient. Below
certain temperature, $\xi_3^{\prime\, 2}$ in
Eq.~(\ref{constantsprime}) becomes negative, and interpretation of
the potential in Eq.~(\ref{Vquad}) as that of a harmonic oscillator,
and the sum of convergent geometric series that led to Eqs.\
(\ref{Iquad}) and (\ref{2eqs}) is no longer justified. Nevertheless,
the expressions in Eqs.\ (\ref{Iquad}) and (\ref{2eqs}) are analytic
functions of $\xi_3^{\prime\, 2}$, than remain meaningful and are
expected to remain valid beyond the range in which they were proven.
Numeric implementation of the Hartree approximation requires special
care in order to pick the relevant rather than spurious solutions of
Eqs.~(\ref{2eqs}). In the limit $T\rightarrow 0$, $R/\xi
'_3\rightarrow i\varphi $, so that the right hand side in the first
of Eqs.~(\ref{2eqs}) does not vanish. Taking the appropriate limits in Eqs.\
(\ref{Iquad}) and (\ref{2eqs}) we obtain that the current in the
Hartree approximation is $I_H(0)=-4\pi
w(\Delta_1^2+\Delta_2^2)\varphi /R\Phi_0$, exactly as
in the absence of fluctuations.

\begin{figure}
\includegraphics[width=\linewidth]{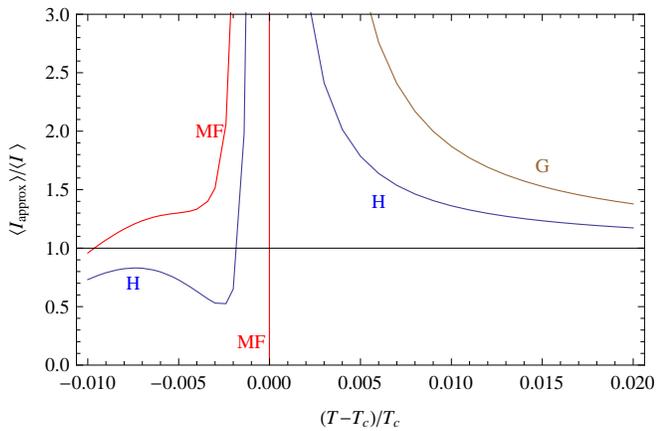}
\caption{\label{approx} Ratio of the current values obtained with
approximated methods to the exact evaluation. Parameters are the
same as in Fig.~\ref{Delta12}, except that here $\varphi =0.25$. G:
Gaussian (quadratic) approximation; H: Hartree approximation; MF:
mean field evaluation (i.e., without fluctuations). The descent of
the MF curve to 0 at the Little--Parks temperature looks vertical in
this scale.}
\end{figure}

We conclude this section with a review of the accuracy of our
evaluations. The accuracy of the ``exact'' evaluation can be
estimated by repeating it with reduced maximum values for $n_\nu$
and $|\ell_\nu|$. We found the largest inaccuracy for large $w$ and
small $T$. In the results presented in Fig.~\ref{Ioftw}, the maximal
inaccuracies are of the order of 10\%. Figure \ref{approx} compares
our approximation methods against the exact evaluation for
$w=10^{-10}{\rm cm}^2$. One can see in the figure that all the
approximation methods are very inaccurate precisely in the most
interesting region, i.e., close to $T_c$. The range of temperatures
where the approximations are inaccurate is larger for smaller
cross-sections $w$. Note also that there exists a range of
temperatures for which the mean field current is larger than our
exact evaluation, meaning that thermal fluctuations {\it inhibit the
current} in this region.

\section{Independence and Asymmetry of the Order Parameters}
The most conspicuous qualitative differences of a two order
parameter system, as compared to a system with a single order
parameter, are {\it independence} and {\it asymmetry}. By
independence we mean that at a given point and time the two order
parameters are not necessarily equal to each other; by asymmetry we
mean that the average values of the order parameters are not necessarily
the same. In this section we investigate the influence of these
properties.

\subsection{Symmetric case}
We start by considering independence while assuming equal
coefficients for both order parameters. For simplicity, we assume
\be a_\nu =\gamma -\alpha \tau \;,\;\; b_\nu =b \;,
\label{symmetric} \ee with $\gamma $, $\alpha$ and $b$ constants. As
an illustration, we may think of a film of a uniform
single-parameter material of thickness $z_0$ with energy density
$-2\alpha \tau |\Delta |^2+b|\Delta |^4+2|{\bm \nabla}\Delta |^2$.
If we decide to denote by $\Delta_1$ (resp.\ $\Delta_2$) the value
of $\Delta$ in the upper (resp.\ lower) half of the film, substitute
the $z$-derivative by a finite difference and average over $z_0$, we
obtain the energy density $-\alpha \tau (|\Delta_1 |^2+|\Delta_2
|^2)+b(|\Delta_1 |^4+|\Delta_2 |^4)/2+|{\bm
\nabla}_{xy}\Delta_1|^2+|{\bm
\nabla}_{xy}\Delta_2|^2+(8/z_0^2)(|\Delta_1 |^2+|\Delta_2
|^2-\Delta_1 \Delta_2^*-\Delta_2 \Delta_1^*)$, with ${\bm
\nabla}_{xy}$ being the component of the gradient in the plane of
the film. One can easily identify that we have recovered the energy
density for two order parameters, with coupling $\gamma =8/z_0^2$.
In the limit $z_0\rightarrow 0,\;\gamma\rightarrow\infty$, and
$\Delta_1$ and $\Delta_2$ are the same; in the opposite extreme,
$\gamma\rightarrow 0$, and $\Delta_1$ and $\Delta_2$ are
independent, while in the general case they are correlated.
Following this analogy, it is very easy to obtain results for the
cases $\gamma\rightarrow 0$ and $\gamma\rightarrow\infty$: the case
$\gamma\rightarrow 0$ is equivalent to that of two single parameter
systems in parallel, and the case $\gamma\rightarrow\infty$ is
equivalent to that of a single parameter system with a doubled cross
section.

\begin{figure}
\includegraphics[width=\linewidth]{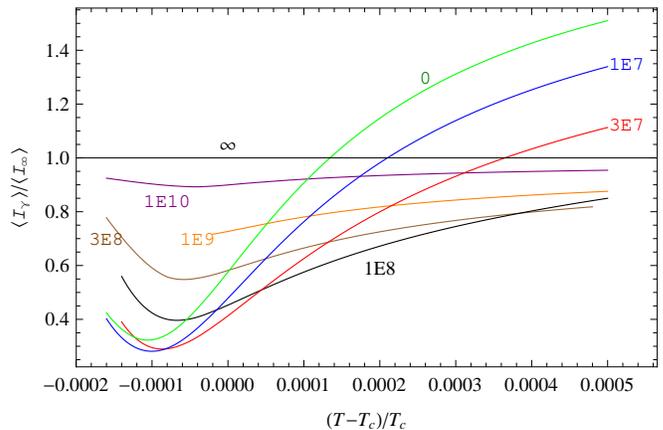}
\caption{\label{independent}Comparison among currents as functions
of the temperature for systems with different coupling $\gamma$, but
with otherwise identical parameters. The value of $\gamma $ is
marked next to every curve in E-notation (e.g. 3E7 denotes $\gamma
=3\times 10^7 {\rm cm}^{-2}$). The other parameters are
$R=10^{-4}\,$cm, $w=10^{-10}\,$cm$^2$, $\alpha =10^{12}\,$cm$^{-2}$,
$b=10^{17}\,$erg$^{-1}$cm$^{-1}$, $k_BT_c=10^{-15}\,$erg, $\varphi
=0.25$. }
\end{figure}

Figure \ref{independent} compares calculated average currents
$\langle I(T)\rangle$ as the parameter $\gamma $ is varied in the
range $0\le\gamma <\infty$, while all the other parameters are
common to all curves. For a facilitated comparison, all the
functions have been divided by $\langle I_\infty (T)\rangle$, the
current obtained for $\gamma\rightarrow\infty$. Figure
\ref{independent} shows the temperature range close to $T_c$; far
below $T_c$ the influence of fluctuations is negligible and all the
curves should coalesce. For every temperature, we note that as
$\gamma $ increases from 0 to $\infty$, $\langle I(T)\rangle$
changes from $\langle I_0(T)\rangle$ to $\langle I_\infty
(T)\rangle$. However, this change is not monotonic: $|\langle
I(T)\rangle|$ initially decreases and after reaching a minimum
increases towards $|\langle I_\infty (T)\rangle|$. The fact that
$|\langle I_0(T)\rangle|<|\langle I_\infty (T)\rangle|$ for
$T\approx T_c$ may look surprising, since $\gamma =0$ means larger
freedom than $\gamma\rightarrow\infty$ and we would therefore expect
larger fluctuations in the former case. We will see in the following
that indeed the order parameters assume larger values for $\gamma
=0$; however, they may be less coordinated, resulting in a smaller
current.

\begin{figure}
\includegraphics[width=\linewidth]{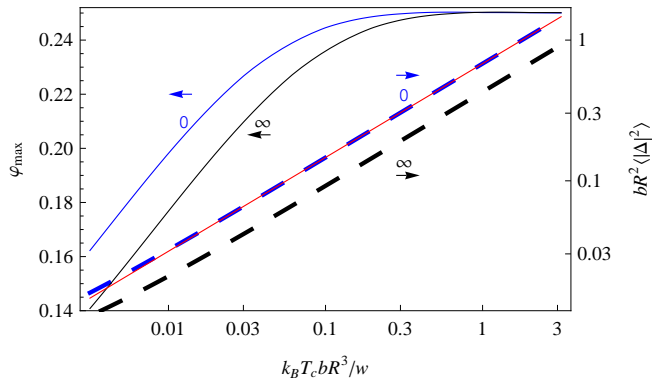}
\caption{\label{fidel} Magnetic flux for which the
fluctuation current is maximal, and value of the order parameters at
$T=T_c$, as functions of $T_c$. $\langle |\Delta|^2\rangle$ was
evaluated at $\varphi =\varphi_{\rm max}$. Each curve is marked by
its value of $\gamma $ and by an arrow that points to the relevant
$y$-axis. The thin straight line (red online) highlights the asymptotic
power dependence of $bR^2\langle |\Delta|^2\rangle$ on
$k_BT_cbR^3/w$. }
\end{figure}

The solid curves in Fig.~\ref{fidel} show the values of $\varphi$
for which the current is maximum at $T=T_c$ for the limiting cases
$\gamma =0$ and $\gamma\rightarrow\infty$. From a dimensional
analysis we find that in the present situation the temperature
enters the operator $H/S$ only through the combination
$k_BT_cbR^3/w$, so that $\varphi_{\rm max}$ is a function of this
quantity. Since the mean field current has its maximum at
$\varphi_{\rm max}>1/4$, it is interesting to note that
$\varphi_{\rm max}$ can be smaller than 1/4. The curve for
$\gamma\rightarrow\infty$ can be inferred from the case $\gamma =0$:
in order to obtain it at a given $T_c$, we have to double the value
of $w$. Since $\varphi_{\rm max}$ is a function of $k_BT_cbR^3/w$,
doubling $w$ is the same as dividing $T_c$ by 2, i.e., at a given
$T_c$, the value of $\varphi_{\rm max}$ for
$\gamma\rightarrow\infty$ is the same that $\varphi_{\rm max}$ for
$\gamma =0$ had at half that temperature. With a logarithmic
$x$-axis, this relation gives a shift of the curve to the right.

The dashed lines in Fig.~\ref{fidel} show the values of $bR^2\langle
|\Delta|^2\rangle$ at $T=T_c$ and $\varphi =\varphi_{\rm max}$,
evaluated by means of Eq.~(\ref{anyp}). Except for $k_BT_c\ll
w/bR^3$ or $k_BT_c\gg w/bR^3$, we obtain $\langle
|\Delta|^2\rangle\approx 0.68
(bR^2)^{-1}(k_BT_cbR^3/w)^{2/3}=0.68(k_B^2T_c^2/bw^2)^{1/3}$ for
$\gamma =0$. For $\gamma\rightarrow\infty$ $\langle
|\Delta|^2\rangle$ is smaller by a factor $2^{2/3}$. The first term
in the Hartree approximation value in Eq.~(\ref{largeR}) is smaller
than the result obtained for $\gamma\rightarrow\infty$ by about 7\%.

\begin{figure}[t]
\includegraphics[width=\linewidth]{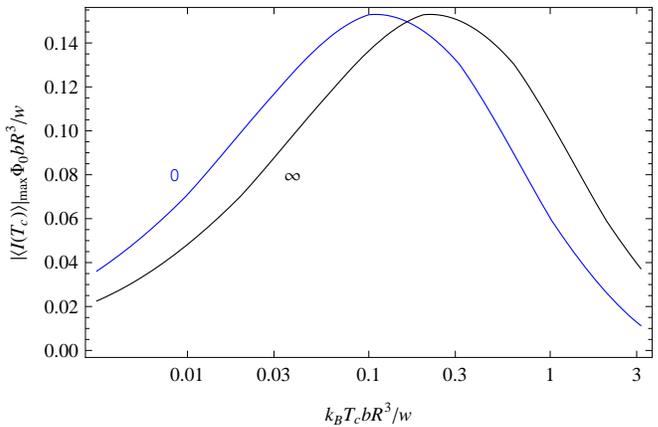}
\caption{\label{paamon} Maximum current at $T=T_c$ as a function of
the scaled $T_c$ for $\gamma =0$ and for
$\gamma\rightarrow\infty$.}
\end{figure}

Figure \ref{paamon} shows the average current evaluated at $T=T_c$
and $\varphi =\varphi_{\rm max}$. As already found in
Ref.~\onlinecite{vOR}, $\langle I(T_c)\rangle$ is not a monotonic
function of $T_c$, but has a maximum instead. As discussed above,
the curve for $\gamma\rightarrow\infty$ is obtained as a shift of
the curve for $\gamma =0$. What we learn from this curve is that for
$k_BT_c<0.163 w/bR^3$ $|\langle I_0(T_c)\rangle |> |\langle
I_\infty(T_c)\rangle |$, meaning that independence of the order
parameters {\it enhances} the fluctuation current, whereas the {\it
opposite} occurs for $k_BT_c>0.163 w/bR^3$.
\begin{figure}
\includegraphics[width=\linewidth]{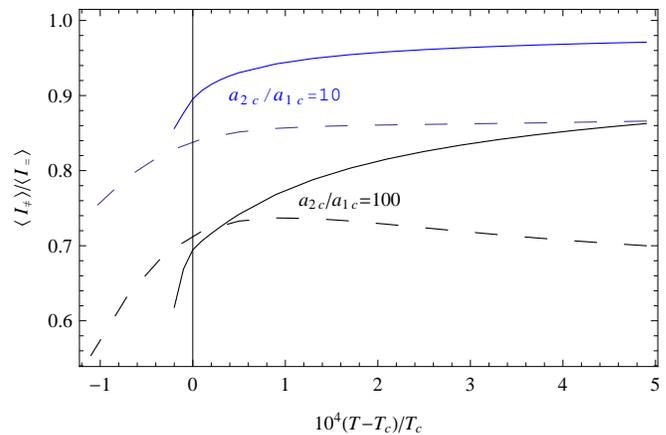}
\caption{\label{diffas} Ratio between the fluctuation currents for
the case $a_{1c}\neq a_{2c}$ (denoted $I_{\neq}$) and the case
$a_{1c} = a_{2c}$ (denoted $I_=$). For the solid lines $b=3\times
10^{15}$erg$^{-1}$cm$^{-1}$ ($k_BT_cbR^3/w=0.03$) and $\gamma
=3\times 10^6$cm$^{-2}$; for the dashed lines
$b=10^{17}$erg$^{-1}$cm$^{-1}$ ($k_BT_cbR^3/w=1$) and $\gamma
=3\times 10^7$cm$^{-2}$. In all cases $\varphi =\varphi_{\rm max}$.
The other parameters are the same as in Fig.~\ref{independent}.}
\end{figure}

\subsection{Asymmetric case}
There are three material parameters that can differ between the
order parameters: $a_{1c}\neq a_{2c}$, $\alpha_1\neq\alpha_2$ and
$b_1\neq b_2$. Since near $T_c$ the usual case is $|\alpha_{1,2}\tau
|, b_{1,2}|\Delta_{1,2}|^2\ll |a_{1c}-a_{2c}|$, we focus on the
influence of the difference between $a_{1c}$ and $a_{2c}$.

Figure \ref{diffas} shows the ratio between the fluctuation currents
for the cases $a_{1c}\neq a_{2c}$ and $a_{1c} = a_{2c}$, while all
the other parameters are kept unchanged. Although the values of
$k_BT_cbR^3/w$ and $\gamma $ do have some influence, the general
trend is that the difference between $a_{1c}$ and $a_{2c}$ inhibits
fluctuation supercurrent in the region $T\approx T_c$, with this
effect being stronger for $T< T_c$.

\section{Conclusions}
Motivated by recent surge in interest in the physics of coupled
condensates in two-band superconductors, we have analyzed the role and
importance of fluctuations in superconducting rings with two order
parameters. We have extended the analysis of the fluctuative regime made
by von Oppen and Riedel\cite{vOR} for the single order
parameter case, based on the Ginzburg--Landau energy functional.
Further, we have made semi-analytic evaluations of the influence of
fluctuations on the persistent current and on the order parameters in the
ring, as functions of temperature, coupling between the order
parameters, and magnetic flux. We have identified the ranges of
parameters where fluctuations inhibit or enhance the persistent
current in the ring, and pointed out the influence of the cross
section of the ring as well as the influence of the freedom of the order parameters to undergo separate fluctuations.
Although the influence of fluctuations is most important close to $T_c$, we have also studied the behavior far from $T_c$, providing a complete picture.

In addition to two-band materials, our findings apply to
artificially made systems of two superimposed rings, as encountered
in experiments that involve annular Josephson
junctions. The present study can also serve as a guideline
for theoretical efforts and interpretations of 
experimental data in systems described by multiple order parameters in the
fluctuative regime, especially in nanothin samples, which are
always effectively multiband due to quantum confinement\cite{nano} and where fluctuations are of outmost
importance.\cite{pogosov}

\begin{acknowledgments}
This research was supported by the Israel Science Foundation, grant
249/10, the Flemish Science Foundation (FWO-Vl), and the ESF network
INSTANS. We are grateful to Andrei Varlamov and Felix von Oppen for
their answers to our enquiries.
\end{acknowledgments}

\end{document}